\documentclass[letterpaper,english,reprint, aps,pra,superscriptaddress]{revtex4-2}
\usepackage[T1]{fontenc}
\usepackage[latin9]{inputenc}
\usepackage{amsmath}
\usepackage{amssymb}
\usepackage{graphicx}

\makeatletter

\pdfpageheight\paperheight
\pdfpagewidth\paperwidth

\makeatother

\usepackage{babel}
\begin{document}
\title{Distance-dependent emission spectrum from two qubits in a strong-coupling regime}

\author{Rongzhen Hu}

\author{JunYan Luo}
\affiliation{Department of Physics, School of Science, Zhejiang University of Science and Technology, Hangzhou 310023, China}

\author{Yiying Yan}\email{yiyingyan@zust.edu.cn}
\affiliation{Department of Physics, School of Science, Zhejiang University of Science and Technology, Hangzhou 310023, China}

\begin{abstract}
We study the emission spectrum of two distant qubits strongly coupled to a waveguide by using the numerical and analytical approaches, which are beyond the Markovian approximation and the rotating-wave approximation (RWA). The numerical approach combines the Dirac-Frenkel time-dependent
variational principle with the multiple Davydov $D_{1}$ ansatz. A transformed RWA (TRWA) treatment and a standard perturbation (SP) are used to analytically calculate the emission spectrum. It is found that the variational approach and the TRWA treatment yield accurate emission spectra of the two distant qubits in certain strong coupling regimes while the SP breaks down. The emission spectrum is found to be asymmetric irrespective of the two-qubit distance and exhibits a single peak, doublet, and multipeaks depending on the two-qubit distance as well as the initial states. In sharply contrast with the single-qubit case, the excited-state populations of the two qubits can ultraslowly decay due to the subradiance even in the presence of a strong qubit-waveguide coupling, which in turn yields ultranarrow emission line. Our results provide insights into the emission spectral features of the two distant qubits in the strong light-matter coupling regime.
\end{abstract}
\date{\today}
\maketitle

\section{Introduction}

Few distant emitters interacting with electromagnetic fields has received much attention, which is fundamentally relevant to the building block of quantum networks~\cite{Samutpraphoot_2020,Ma_2023} and gates~\cite{Cho_2005,Zheng_2009} and also provides a test bed to study a variety of phenomena such as quantum interference~\cite{Wiegner_2011,Lee_2023}, collective emission such as superradiance and subradiance~\cite{Raiford_1974,Fu_1992,Trung,Macovei_2003,Zeeb_2015,Reimann_2015,Bhatti_2018,Pleinert_2018,Guimond_2019,Berman_2020,Solano_2023, Richter_2023}, photon-mediated interaction~\cite{Milonn,Milonn,Rist_2008,van_Loo_2013,D_az_Camacho_2015,Zhao_2017} and quantum entanglement~\cite{Ficek_2002,Lazarou_2008,Cao_2009,L__2009,Nihira_2009,Almutairi_2011,Wang_2020}. Particularly, the literature has illustrated that the nonMarkovianity of the two distant emitters arises from delay-feedback effects due to the field propagating between the distant emitters~\cite{Sinha_2020_prl}, which strongly influences the collective dynamics of the emitters as well as the spontaneous emission.

Recently, the two-atom problem has been renewed in terms of artificial atoms coupled to a waveguide, which allows the access to the coupling regimes from weak to ultrastrong light-matter interaction~\cite{Frisk_Kockum_2019,Forn_D_az_2019}. For instance, Ref.~\cite{Wen_2019} reports the observation of a large collective Lamb shift of two distant superconducting artificial atoms.
With the rotating-wave approximation (RWA), the investigation has revealed the delay-induced nonMarkovian features of the spontaneous emission spectrum of two distant qubits weakly coupled to a one-dimensional waveguide, that is, the linewidth broadening beyond standard supperradiance or narrow Fano-resonance-like peaks~\cite{Sinha_2020}. In the ultrastrong-coupling regime, the photon-mediated interaction between the two distant qubits and
the qubit frequency renormalization are found to be significant and play a crucial role in the collective dynamics of the two-qubit system~\cite{Gonz_lez_Guti_rrez_2021}. However, the spontaneous emission spectrum from two distant emitters has not yet been explored in the strong light-matter coupling regime. This acquires a nonMarkovian and nonRWA approach that properly takes into account strong-coupling effects.

One way to theoretically calculate the emission spectrum of two distant qubits is the time-dependent variational approach equipped with the multiple Davydov ansatz~\cite{Zhao_2023}. This approach has been applied to a variety of models ranging from the quantum Rabi model~\cite{Werther_2018} to Holstein model~\cite{Chen_2017,Werther_2020} and has been benchmarked with other numerical methods such as time-dependent numerical renormaliztion group~\cite{Orth_2010,Deng_2016}, quasi-adiabatic path integral~\cite{Fujihashi_2017}, and hierarchical equation of motion~\cite{Wang_2016,Yan_2020,Chen_2023}. It turns out that the variational approach is capable of describing nonMarkovian dynamics of open quantum systems such as the spin-boson model and its variants in the strong system-reservoir coupling regime. An advantage of the variational approach over the master equation is that it retains both the reduced dynamics and the field dynamics. It is therefore feasible to calculate emission spectrum by passing the two-time correlation function and quantum regression theory~\cite{Yan_2020}. Nevertheless, the performance of the variational approach should be further explored.

In this paper, we employ the time-dependent variational approach and analytical methods to study the emission spectra of  two distant qubits strongly coupled to a one-dimensional waveguide beyond the weak-coupling regime and without using the RWA and Markovian approximation. The variational method combines the Dirac-Frenkel time-dependent variational principle~\cite{frenkel} with the multiple Davydov $D_1$ (multi-$D_{1}$) ansatz~\cite{Deng_2016}, which is found to be able to provide accurate results in certain strong-coupling regime. We also attempt to calculate emission spectrum with two analytical methods. One is based on the resolvent-operator formalism~\cite{Cohen} and a transformed RWA (TRWA) Hamiltonian constructed from a unitary transformation that resembles the polaron transformation~\cite{Cao_2009,Zheng_2004}. The other is the standard perturbation (SP) based on the resolvent-operator formalism and original Hamiltonian. Comparisons between the variational results and analytical results confirm the validity of the TRWA treatment while the SP treatment completely breaks down. By using the variational approach and TRWA method, we illustrate that the emission spectrum is generally asymmetric and has a variety of lineshapes, which can be single-peaked, vacuum-Rabi-splitting-like doublet, and complicated multipeaked depending on the distance and initial states. Under certain conditions, ultranarrow emission lines can be observed in the spectrum, indicating the subradiance. The present study reveals the emission spectral features of the two distant qubits in a strong light-matter coupling regime.

The rest of paper is organized as follows. In Sec.~\ref{sec:MM}, we introduce the Hamiltonian and present both the numerical and analytical treatments for the emission spectrum. In Sec.~\ref{sec:results}, we present the numerical results of the emission spectra and discuss the spectral features. In Sec.~\ref{sec:con}, the conclusions are drawn. Some technique details are presented in Appendices.

\section{Model and methodologies}\label{sec:MM}

We consider that two distant qubits are strongly coupled to a one-dimensional waveguide (reservoir), which is described by the Hamiltonian $(\hbar=1)$
\begin{eqnarray}
H & = & \frac{\omega_{0}}{2}\sum_{j=1}^{2}\sigma_{j}^{z}+\sum_{k}\omega_{k}b_{k}^{\dagger}b_{k}\nonumber \\
 &  & +\sum_{j=1}^{2}\frac{\sigma_{j}^{x}}{2}\sum_{k}\lambda_{k}\left(b_{k}e^{-ikx_{j}}+b_{k}^{\dagger}e^{ikx_{j}}\right),\label{eq:Ham}
\end{eqnarray}
where $\omega_{0}$ is the transition frequency of the qubit, $\sigma_{j}^{\mu}$$(\mu=x,y,z)$
denotes the Pauli matrix for the $j$th qubit. $x_{j}$ is the coordinate
of the $j$th qubit. $b_{k}$ ($b_{k}^{\dagger}$) is annihilation (creation)
operator of the $k$th bosonic mode with frequency $\omega_{k}$ of
the reservoir. $\lambda_{k}$ is the coupling strength between
the $k$th mode and the qubit. In this work, we assume $\lambda_{k}=\lambda_{-k}$
and a linear dispersion relation $\omega_{k}=v_{g}|k|$,
where $v_{g}$ is the propagating velocity of the photon in the waveguide. The wavenumber $k<0$ and $k>0$ are referred to as the left- and right-propagating field modes in the waveguide, respectively.

The dissipation of the waveguide is assumed to be characterized by the Ohmic spectral density function
\begin{equation}
J(\omega)=\sum_{k}\lambda_{k}^{2}\delta(\omega_{k}-\omega)=2\alpha\omega\Theta(\omega_{c}-\omega),
\end{equation}
where $\alpha$ is a dimensionless coupling constant, $\Theta(\cdot)$
is the Heaviside function, and $\omega_{c}$ is the cut-off frequency. In the following, we use numerical and analytical methods to study the spontaneous emission of the two-qubit system.

\subsection{Dirac-Frenkel time-dependent variational principle and multi-$D_1$ ansatz}

In this section, we use the numerical approach that combines the Dirac-Frenkel time-dependent variational
principle with the multi-$D_{1}$ ansatz to study the spontaneous emission. This approach is feasible to solve the time-dependent Schr\"{o}dinger
equation $i\frac{d}{dt}|\tilde{\psi}(t)\rangle=\tilde{H}(t)|\tilde{\psi}(t)\rangle$
in the interaction picture governed by the bath Hamiltonian $H_{R}=\sum_{k}\omega_{k}b_{k}^{\dagger}b_{k}$,
where
\begin{equation}
\tilde{H}(t)=\frac{\omega_{0}}{2}\sum_{j=1}^{2}\sigma_{j}^{z}+\sum_{j=1}^{2}\frac{\sigma_{j}^{x}}{2}\sum_{k}\lambda_{k}\left(b_{k}e^{-ikx_{j}-i\omega_{k}t}+{\rm h.c.}\right).
\end{equation}
The Dirac-Frenkel time-dependent variational principle states that
the optimal solution to the time-dependent Schr\"{o}dinger equation can be found
via~\cite{frenkel}
\begin{equation}
\langle\delta\tilde{\psi}(t)|i\partial_{t}-\tilde{H}(t)|\tilde{\psi}(t)\rangle=0,\label{eq:DFTDVP}
\end{equation}
where $|\tilde{\psi}(t)\rangle$ denotes a trial state and $\langle\delta\tilde{\psi}(t)|$
is the variation of the adjoint state of the trial state. Since the model under study in this work is a variant of the spin-boson model, we use
the multi-$D_1$ ansatz, which has been found to be powerful in the spin-boson problem and takes the form~\cite{Deng_2016}
\begin{eqnarray}
|D_{1}^{M}(t)\rangle & = & \sum_{j=1}^{4}\sum_{n=1}^{M}A_{nj}|\phi_{j}\rangle|f_{nj}\rangle,
\end{eqnarray}
where $M$ is the multiplicity, $|\phi_{j}\rangle\in\{|+\rangle|+\rangle,|+\rangle|-\rangle,|-\rangle|+\rangle,|-\rangle|-\rangle|\sigma^{x}|\pm\rangle=\pm|\pm\rangle\}$
are the bases for the two-qubit system. $|f_{nj}\rangle$ are the multimode coherent state used for the bosonic modes,
\begin{equation}
|f_{nj}\rangle=\exp\left[\sum_{k}(f_{njk}b_{k}^{\dagger}-{\rm h.c.})\right]|0\rangle,
\end{equation}
where $|0\rangle$ is the multimode vacuum state of the reservoir.
Supposing that the truncation numbers of the left- and
right-propagating modes in the multi-$D_{1}$ ansatz are identical and given by $N_{b}$, we have introduced
totally $4M(2N_{b}+1)$ time-dependent variational parameters: $A_{nj}$
and $f_{njk}$. The former represents the probability amplitude while
the latter represents the displacement of the $k$th mode. One readily derives the equations
of motion for these variational parameters by substituting
the ansatz into Eq.~\eqref{eq:DFTDVP}, which yields
\begin{widetext}
\begin{equation}
i\langle\phi_{j}|\langle f_{mj}|\dot{D}_{1}^{M}(t)\rangle=\langle\phi_{j}|\langle f_{mj}|\tilde{H}(t)|D_{1}^{M}(t)\rangle, \label{eq:eom1}
\end{equation}
\begin{equation}
i\sum_{j=1}^{4}A_{mj}^{\ast}\langle\phi_{j}|\langle f_{mj}|b_{q}|\dot{D}_{1}^{M}(t)\rangle=\sum_{j=1}^{4}A_{mj}^{\ast}\langle\phi_{j}|\langle f_{mj}|b_{q}\tilde{H}(t)|D_{1}^{M}(t)\rangle. \label{eq:eom2}
\end{equation}
\end{widetext}
This is a set of nonlinear differential equations, which can be solved by using the Runge-Kutta method. We present
the explicit form of the equations of motion and state the numerical implementation of the variational method in the Appendix~\ref{app:EOM}.

To perform numerical simulation, we use a finite truncated number of bath
modes, $2N_{b}$, which can be derived from a linear discretization
of the spectral density. We divide the frequency domain {[}0,$\omega_{c}${]}
into $N_{b}$ equal intervals $[x_{n-1},x_{n}]$ with $x_{n}=n\omega_{c}/N_{b}$
$(n=0,1,2,...,N_{b}).$ The coupling constants and frequencies for the right-propagating modes ($k>0$) are determined by
\begin{equation}
\lambda_{k_{n}}^{2}=\frac{1}{2}\int_{x_{n-1}}^{x_{n}}J(\omega)d\omega,
\end{equation}
\begin{equation}
\omega_{k_{n}}=\frac{1}{2}\lambda_{k_{n}}^{-2}\int_{x_{n-1}}^{x_{n}}\omega J(\omega)d\omega.
\end{equation}
where we have assumed that the left- and right-propagating modes contribute equally to the spectral density function and $1/2$ is used to cancel out the contribution from the left-propagating
modes. The frequencies and coupling constants of the left-propagating modes can be obtained from the relations: $\omega_{k}=\omega_{-k}$ and
$\lambda_{k}=\lambda_{-k}$. The wavenumber is then specified via the relation $k_{n}=\pm\omega_{k_{n}}/v_{g}$. Throughout this work, we use $N_b=300$ in the simulation, which is sufficient to yield convergent results when the final evolution time $t\leq300\omega_{0}^{-1}$ and $M$ is large enough.

We are interested in the spontaneous emission process of the two-qubit system, thus
the initial state of the whole system is chosen as $|\psi(0)\rangle_{S}\otimes|0\rangle,$
where $|\psi(0)\rangle_{S}$ is an initial state of the two-qubit
system and the reservoir is initially in the vacuum state. In this work, we mainly consider three kinds of initial states of the two-qubit system: $\Psi_0=|eg\rangle$ and $\Psi_{\pm}=(|eg\rangle\pm|ge\rangle)/\sqrt{2}$. $\Psi_{0}$ is a factorized state where the first qubit is in the excited state $|e\rangle$ and the second qubit is in the ground state $|g\rangle$. $\Psi_{\pm}$ are the symmetric and antisysmetric correlated (entangled) state, respectively.

On numerically solving the equations of motion, we can obtain both reduced dynamics of the qubits and the field dynamics. The excited-state population of the $j$th qubit can be calculated as
\begin{equation}
  P_{j}^{e}(t)=\langle D^{M}_{1}(t)|\sigma_j^{+}\sigma_j^{-}|D^{M}_{1}(t)\rangle,\quad (j=1,2)
\end{equation}
where
\begin{equation}
\sigma_{j}^{\pm}=(\sigma_{j}^{x}\pm i\sigma_{j}^{y})/2.
\end{equation}
For the field, we are interested in the emission spectrum, which is defined as the number of photon with frequency $\omega_{k}$ emitted into the reservoir at time $t$. The spontaneous emission spectrum is thus given by
\begin{equation}
N(\omega_{k},t)=N(k,t)+N(-k,t),
\end{equation}
where
\begin{eqnarray}
N(k,t) & = & \langle D_{1}^{M}(t)|b_{k}^{\dagger}b_{k}|D_{1}^{M}(t)\rangle\nonumber \\
 & = & \sum_{j=1}^{4}\sum_{m,n=1}^{M}A_{mj}^{\ast}f_{mjk}^{\ast}\langle f_{mj}|f_{nj}\rangle f_{njk}A_{nj},
\end{eqnarray}
is the photon number at the $k$th mode at given time $t$. $\langle f_{mj}|f_{nl}\rangle$
is the overlap between the coherent states and is given by
\begin{equation}
\langle f_{mj}|f_{nl}\rangle=\exp\left[\sum_{k}\left(f_{mjk}^{\ast}f_{nlk}-\frac{|f_{mjk}|^{2}+|f_{nlk}|^{2}}{2}\right)\right].
\end{equation}
The emission spectrum defined by $N(\omega_k,t)$ is generally time
dependent. Nevertheless, the steady-state spectrum can be obtained
in the long-time limit, i.e.,
\begin{equation}
N(\omega_{k})=\lim_{t\rightarrow\infty}[N(k,t)+N(-k,t)]. \label{eq:Nwk}
\end{equation}
In simulation, we propagate the equations of motion for the variational parameters to a finial time $t=300\omega_0^{-1}$, which is sufficient to obtain steady-state spectra in most cases. The obtained spectra are referred to as the multi-$D_{1}$ results.

To measure the accuracy of the variational results, we calculate the
scaled squared norm of the deviation vector~\cite{Martinazzo_2020},
\begin{eqnarray}
\sigma^{2}(t) & = & |[i\partial_{t}-\tilde{H}(t)]|D_{1}^{M}(t)\rangle|^{2}/\omega_{0}^{2}\nonumber \\
 & = & \omega_{0}^{-2}\left[\langle D_{1}^{M}(t)|\tilde{H}^{2}(t)|D_{1}^{M}(t)\rangle-\langle\dot{D}_{1}^{M}(t)|\dot{D}_{1}^{M}(t)\rangle\right].\nonumber\\
\end{eqnarray}
The detailed calculation and behaviors of $\sigma^{2}(t)$ with the variation of $t$ for the two-qubit spin-boson model are presented in the Appendix~\ref{app:EOM}. We find that the upper bound of the magnitude of $\sigma^2(t)$ in the interval $[0,300\omega_0^{-1}]$ is of order $10^{-3}$ or $10^{-2}$ when $\alpha\leq0.1$.
This is sufficient to guarantee the accuracy of the variational results according to our previous work~\cite{Yan_2020}.
When $\alpha>0.1$, it turns out that the variational method is accurate in short-time dynamics but becomes less reliable in the long-time dynamics because of the increase in the error.

\subsection{Analytical theory for spontaneous emission spectrum}

In this section, we use two approximate approaches to analytically calculate the emission spectra with the resolvent-operator formalism~\cite{Cohen}.
In the first approach, we derive an effective Hamiltonian in a transformed frame and then combine it with the resolvent-operator formalism to evaluate the transition amplitude associated with the spontaneous emission process, which can be used to calculate the photon number at the $k$th mode in the long-time limit, $N(k)$. The steady spectrum is then obtained via Eq.~\eqref{eq:Nwk}. This treatment is referred to as the TRWA method. The second approach is similar to the first one but we use the original Hamiltonian, which is referred to as SP.

To go beyond the weak-coupling regime, we apply a polaron-like unitary
transformation to Eq.~\eqref{eq:Ham}~\cite{Cao_2009,Zheng_2004}
\begin{equation}
  H^{\prime}=e^{S}He^{-S}
\end{equation}
with the generator given by
\begin{equation}
S=\sum_{j=1}^{2}\sigma_{j}^{x}\sum_{k}\frac{\lambda_{k}}{2\omega_{k}}\xi_{k}\left(b_{k}^{\dagger}e^{ikx_{j}}-b_{k}e^{-ikx_{j}}\right),
\end{equation}
where $\xi_{k}$ are determined by requiring
the first-order qubit-reservoir coupling to take the rotating-wave
form in the transformed frame. Neglecting the higher-order qubit-reservoir
coupling terms, we construct an effective Hamiltonian from the transformed Hamiltonian,
\begin{equation}
H^{\prime}\approx H_{0}^{\prime}+H_{1}^{\prime},
\end{equation}
\begin{equation}
H_{0}^{\prime}=\frac{1}{2}\eta\omega_{0}\sum_{j=1}^{2}\sigma_{j}^{z}+\sum_{k}\omega_{k}b_{k}^{\dagger}b_{k}+\sum_{k}\frac{\lambda_{k}^{2}}{2\omega_{k}}(\xi_{k}^{2}-2\xi_{k}),
\end{equation}
\begin{equation}
H_{1}^{\prime}=V_{c}\sigma_{1}^{x}\sigma_{2}^{x}+\sum_{j=1}^{2}\sum_{k}\tilde{\lambda}_{k}\left(\sigma_{j}^{+}b_{k}e^{-ikx_{j}}+\sigma_{j}^{-}b_{k}^{\dagger}e^{ikx_{j}}\right),
\end{equation}
where
\begin{equation}
\eta=\exp\left[-\frac{1}{2}\int_{0}^{\omega_c}\frac{J(x)dx}{(x+\eta\omega_0)^2}\right],
\end{equation}
\begin{equation}
V_{c}=-\int_{0}^{\omega_c}\frac{J(x)(x+2\eta\omega_0)}{2(x+\eta\omega_0)}\cos\left(\frac{xd}{v_g}\right)dx,
\end{equation}
\begin{equation}
  \xi_{k}=\frac{\omega_k}{\omega_k+\eta\omega_0},
\end{equation}
\begin{equation}
d=x_{1}-x_{2},
\end{equation}
\begin{equation}
\tilde{\lambda}_{k}=\frac{\lambda_{k}\eta\omega_{0}}{\eta\omega_{0}+\omega_{k}}.
\end{equation}
This effective Hamiltonian is named as TRWA Hamiltonian.
The detailed derivation of the TRWA Hamiltonian is presented in the Appendix~\ref{app:Heff}.
In the TRWA Hamiltonian, it is worthwhile to note the important consequences of the qubit-reservoir coupling: the renormalization of the transition frequencies ($\eta\omega_0$) and a reservoir induced dipole-dipole coupling ($V_{c}\sigma_{1}^{x}\sigma_{2}^{x}$). These effects have been studied in Refs.~\cite{McCutcheon_2010,Gonz_lez_Guti_rrez_2021} with similar approaches in the strong-coupling regime.

Combining the effective Hamiltonian with the resolvent-operator formalism, we calculate the photon number at the $k$th mode in the long-time limit for the three kinds of the two-qubit initial states when the reservoir is initially in the vacuum state. The detailed calculation is given in the Appendix~\ref{app:sw}.
When the initial state of the two qubits is the factorized state $\Psi_{0}$, we find
\begin{equation}\label{eq:swtrwa1}
N(k) =\tilde{\lambda}_{k}^{2}\left|\frac{\tilde{A}\left(\tilde{\omega}_k\right)+e^{-ikd}\tilde{B}\left(\tilde{\omega}_k\right)}{\tilde{A}^{2}\left(\tilde{\omega}_k\right)-\tilde{B}^{2}\left(\tilde{\omega}_k\right)}+\frac{1}{2\eta\omega_{0}}\right|^{2},
\end{equation}
where
\begin{equation}
  \tilde{\omega}_k=\omega_{k}-\frac{V_{c}^{2}}{2\eta\omega_{0}}
\end{equation}
\begin{equation}\label{eq:Awt}
\tilde{A}(\omega)=\omega-\eta\omega_{0}-\tilde{\Delta}\left(\omega,0\right)+i\tilde{\Gamma}\left(\omega,0\right),
\end{equation}
\begin{equation}\label{eq:Bwt}
\tilde{B}(\omega)=V_{c}+\tilde{\Delta}\left(\omega,d\right)-i\tilde{\Gamma}\left(\omega,d\right),
\end{equation}
\begin{equation}
\tilde{\Delta}\left(\omega,d\right)=P\int_{0}^{\omega_{c}}\frac{J(x)\cos(xd/v_{g})}{\omega-x}\left(\frac{\eta\omega_{0}}{x+\eta\omega_{0}}\right)^{2}dx,
\end{equation}
\begin{equation}
\tilde{\Gamma}(\omega,d)=\pi\left(\frac{\eta\omega_{0}}{\omega+\eta\omega_{0}}\right)^{2}J(\omega)\cos\left(\frac{\omega d}{v_{g}}\right).
\end{equation}
When the two-qubit initial state is $\Psi_{\pm}$,
the photon number at the $k$th mode is found to be given by
\begin{equation}\label{eq:swtrwa2}
N(k) =\tilde{\lambda}_{k}^{2}\left|\frac{1\pm e^{-ikd}}{\sqrt{2}}\right|^{2}\left|\frac{1}{\tilde{A}\left(\tilde{\omega}_{k}\right)\mp\tilde{B}\left(\tilde{\omega}_{k}\right)}+\frac{1}{2\eta\omega_{0}}\right|^{2}.
\end{equation}
Interestingly, it seems that Eq.~\eqref{eq:swtrwa2} is much simpler than Eq.~\eqref{eq:swtrwa1}, suggesting that the spontaneous emission from the correlated states may play a more fundamental role than that from the factorized state. In the following, the spectra obtained based on Eqs.~\eqref{eq:swtrwa1} and~\eqref{eq:swtrwa2} are referred to as the TRWA results.

\begin{figure*}
  \includegraphics[width=2\columnwidth]{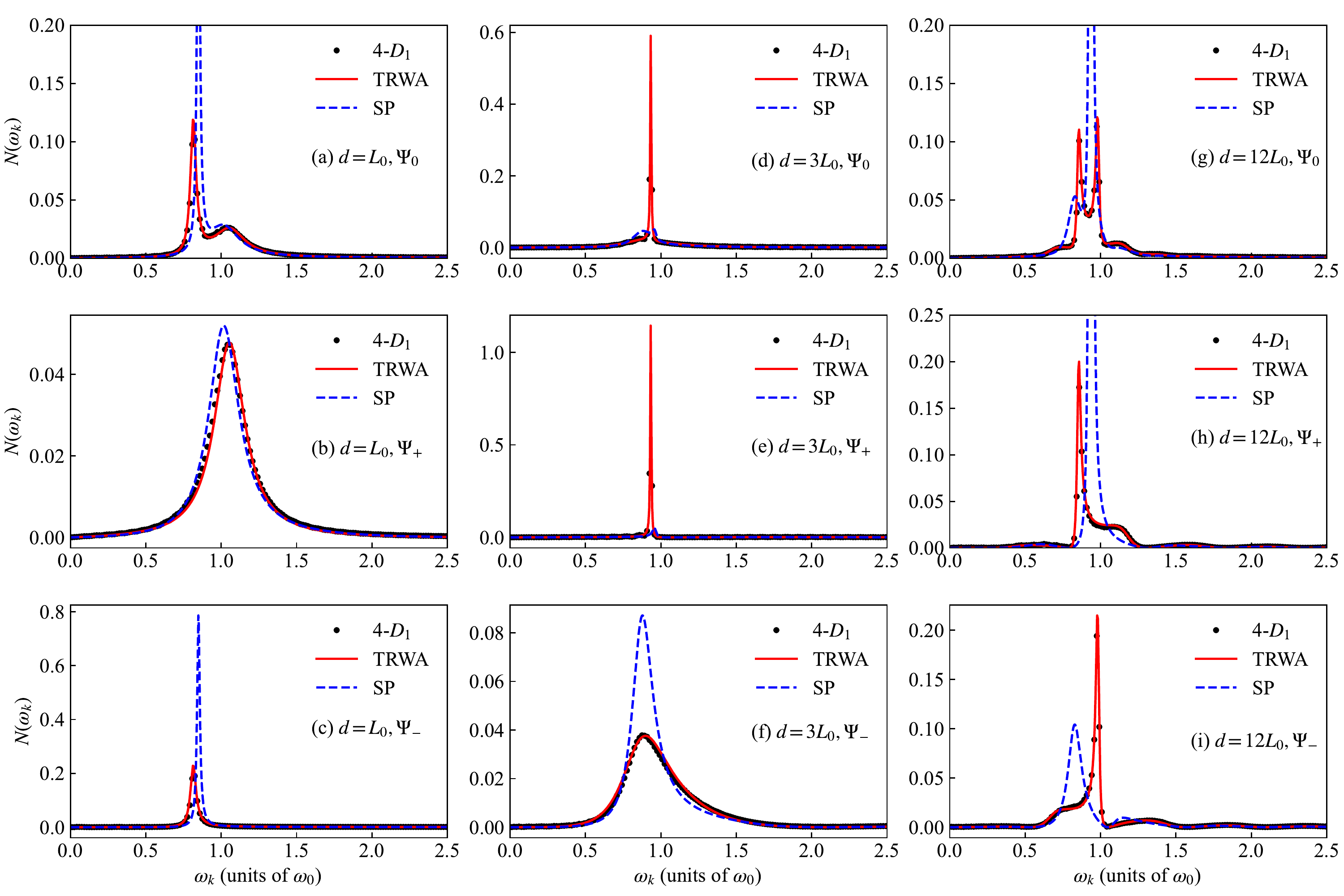}
  \caption{Emission spectra calculated by the three methods for $\alpha=0.05$, the three values of $d$, and the three kinds of initial states. The multiplicity $M$ is indicated in the legends.}\label{fig1}
\end{figure*}

To examine the improvement of the variational and TRWA method with respect to the SP, we further calculate the steady emission spectrum with the resolvent-operator formalism and the original Hamiltonian. The calculation details are presented in the Appendix~\ref{app:sw}. When
the two-qubit initial state is $\Psi_0$, the photon number at the $k$th mode
is given by
\begin{equation}\label{eq:swsp1}
N(k)=\frac{\lambda_{k}^{2}}{4}\left|\frac{A(\omega_{k}^\prime)+e^{-ikd}B(\omega_{k}^\prime)}{A^{2}(\omega_{k}^\prime)-B^{2}(\omega_{k}^\prime)}\right|^{2}
\end{equation}
where
\begin{equation}
  \omega_{k}^\prime=\omega_k+2\Delta(-\omega_{0},0)
\end{equation}
\begin{eqnarray}
A(\omega)  & = &  \omega-\omega_{0}-\Delta(\omega,0)-\Delta(\omega-2\omega_{0},0)\nonumber\\
&  &+i\Gamma(\omega,0)+i\Gamma(\omega-2\omega_{0},0), \label{eq:Aw}
\end{eqnarray}
\begin{equation}
B(\omega)=\Delta(\omega,d)+\Delta(\omega-2\omega_{0},d)-i\Gamma(\omega,d)-i\Gamma(\omega-2\omega_{0},d), \label{eq:Bw}
\end{equation}

\begin{equation}
\Delta(\omega,d)=P\int_{0}^{\omega_{c}}\frac{J(x)\cos(xd/v_{g})}{4(\omega-x)}dx,
\end{equation}
\begin{equation}
\Gamma(\omega,d)=\frac{\pi}{4}J(\omega)\cos(\omega d/v_{g}).
\end{equation}
When the two-qubit initial state is
$\Psi_{\pm}$,
the photon number at the $k$th mode is given by
\begin{equation}\label{eq:swsp2}
N(k)=\frac{\lambda_{k}^{2}}{4}\left|\frac{\frac{1}{\sqrt{2}}(1\pm e^{-ikd})}{A(\omega_{k}^\prime)\mp B(\omega_{k}^\prime)}\right|^{2}.
\end{equation}
Hereafter, the spectra calculated based on Eqs.~\eqref{eq:swsp1} and~\eqref{eq:swsp2} are referred to as the SP results. In the following we will address the validity of the analytical results in comparison with the variational results.

\begin{figure*}
  \includegraphics[width=2\columnwidth]{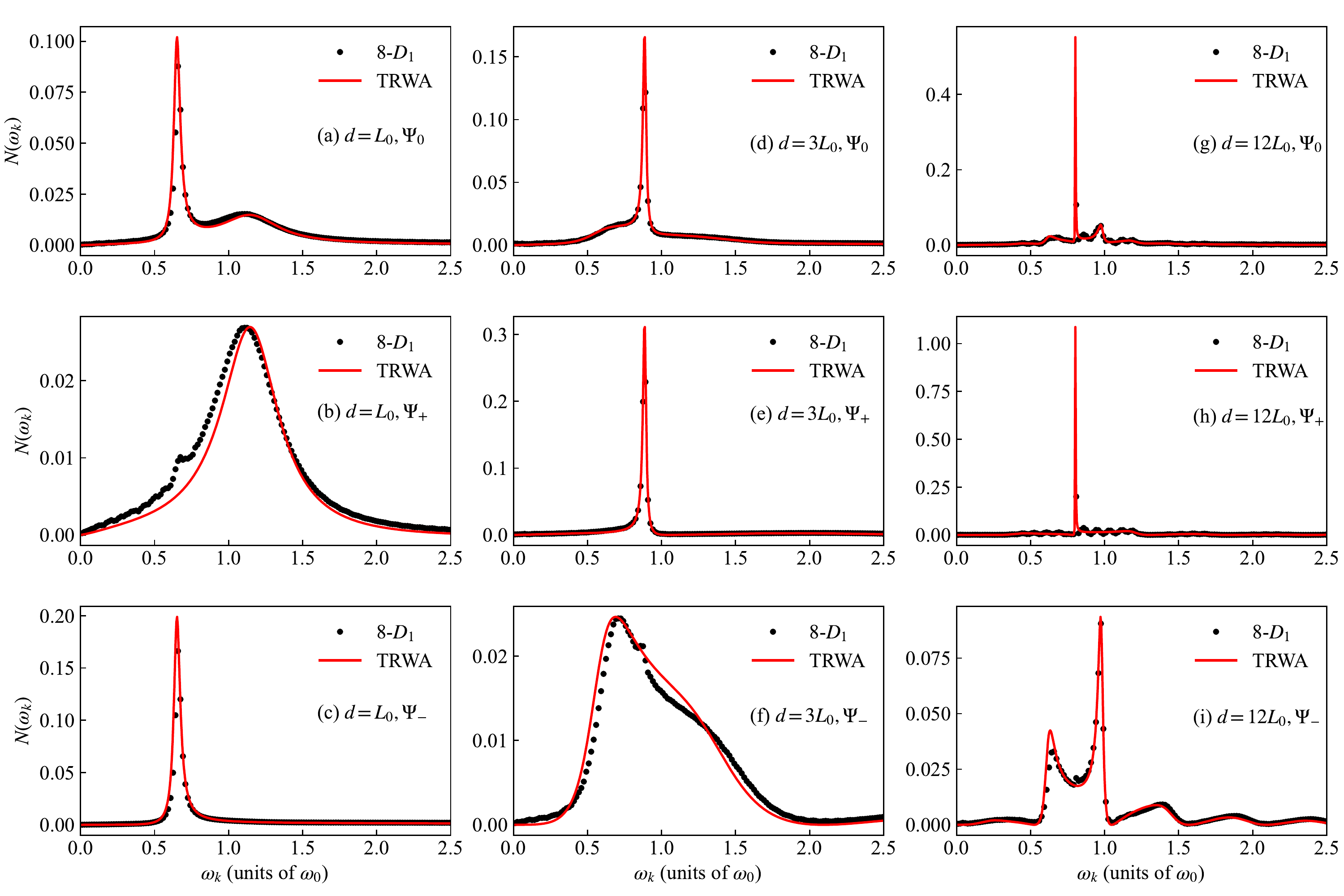}
  \caption{Emission spectra calculated by the variational method and the TRWA method for $\alpha=0.1$, the three values of $d$, and the three kinds of initial states.}\label{fig2}
\end{figure*}

\section{Numerical results and discussions}\label{sec:results}

\begin{figure*}
  \includegraphics[width=\columnwidth]{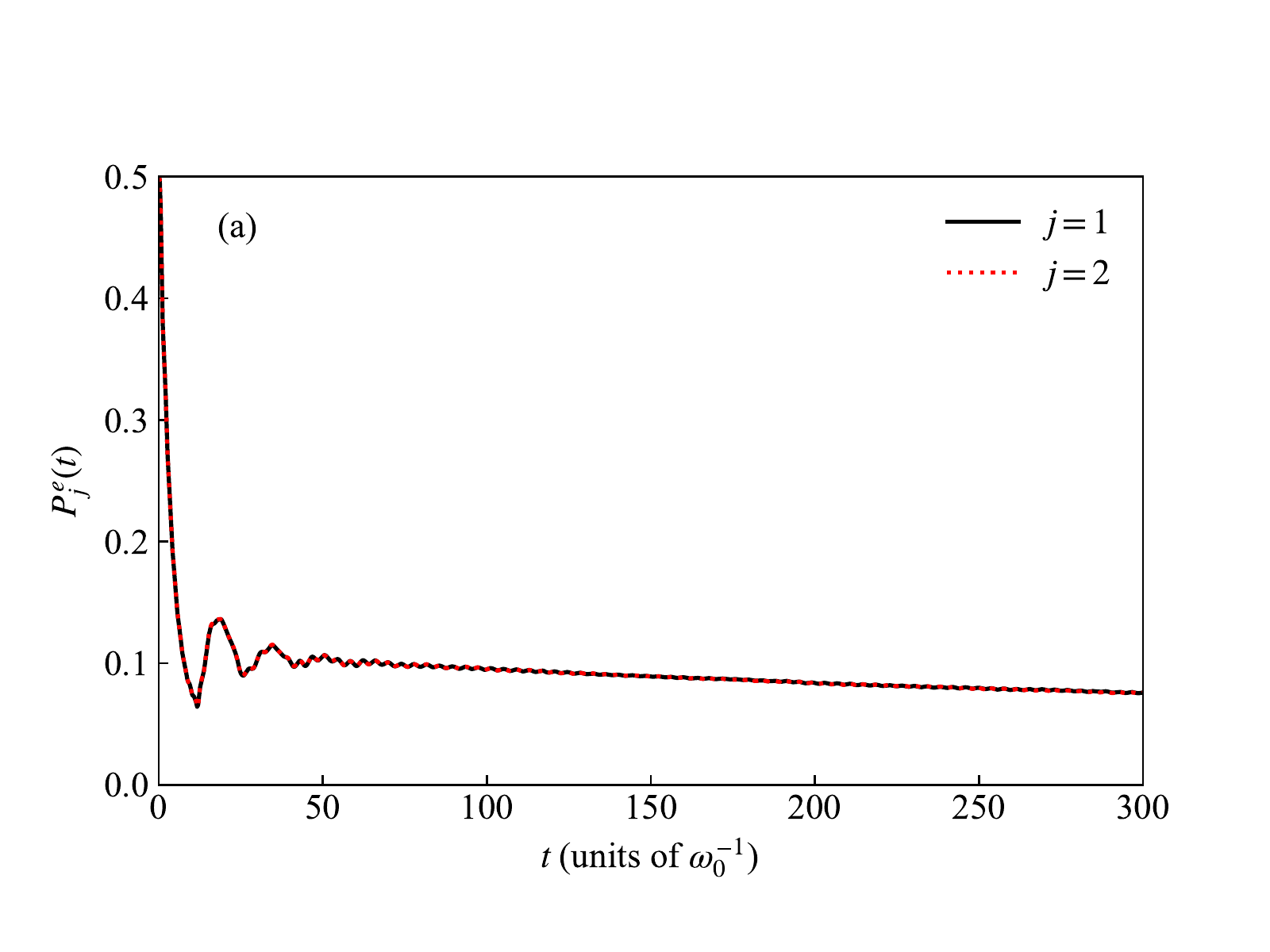}
  \includegraphics[width=\columnwidth]{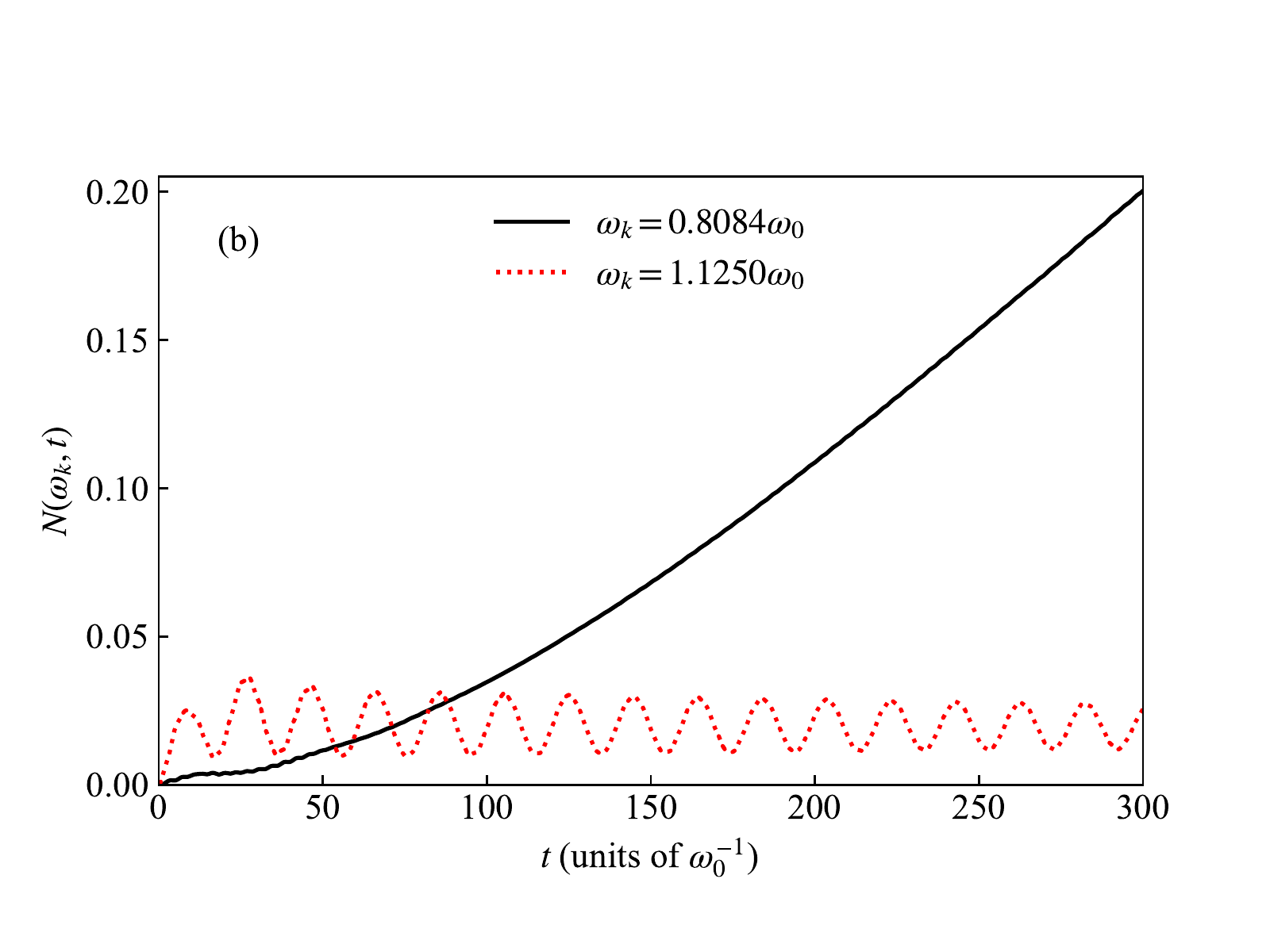}
  \caption{(a) Excited-state population of the $j$th qubit $P_{j}^{e}(t)$ as a function $t$ obtained by the variational method. (b) Photon number $N(\omega_k,t)$ as a function of $t$ obtained by the variational method for the two values of $\omega_k$. The parameters are set as $\alpha=0.1$ and $d=12L_0$. The initial state is the symmetric correlated state $\Psi_{+}$.}\label{fig3}
\end{figure*}

In this section, we calculate the emission spectra by using the three methods: the variational method, TRWA, and SP. We make comparison between the multi-$D_1$ results and the analytical ones. This helps to clarify not only the emission spectral features but also the validity of the analytical treatments. Throughout this work, we set the cut-off frequency $\omega_c=5\omega_0$ and define $L_{0}\equiv v_g/\omega_0$ as a unit of distance.

To begin with, let us address the consistency among the three approaches.
Figure~\ref{fig1} shows the emission spectra calculated by the three methods for $\alpha=0.05$ and for the three kinds of initial states and the three values of the two-qubit distance $d$.
It is evident that the multi-$D_1$ results and the TRWA ones agree perfectly with each other, suggesting the validity of the latter. The SP results are found to be inconsistent with the multi-$D_1$ spectra. Particularly, for larger values of $d$, the SP results are strikingly different from the multi-$D_1$ results. The present results suggest that the variational method and the TRWA treatment are applicable to a strong-coupling regime where the SP treatment breaks down. Nevertheless, we should point out that the SP treatment and the TRWA treatment slightly differ from each other when $\alpha=0.01$ and become indistinguishable from each other when $\alpha=0.001$, that is, the two analytical approaches coincide in the weak-coupling regime.

To further examine the consistency between the variational method and the TRWA one, we employ the two methods to calculate the emission spectra for $\alpha=0.1$ and for the three kinds of the initial states and the three values of $d$, which are shown in Fig.~\ref{fig2}. We see that the TRWA results are satisfactorily accurate in comparison with the multi-$D_{1}$ results, confirming the validity of the TRWA method. It is worthwhile to note that in Figs.~\ref{fig2}(g) and~\ref{fig2}(h), there are more peaks in the multi-$D_{1}$ spectra than in the TRWA spectra. This difference can be attributed to the fact that the multi-$D_1$ spectra shown are not actually in the steady state, that is, we just propagate the equations of motion of the variational approach to the final time $t=300\omega_0^{-1}$ and use $N(\omega_k,t)|_{t=300\omega_0^{-1}}$ to approximate the steady-state spectrum. This approximation may not be justified when a subradiant state with an utlraslow decay is encountered. We will discuss such phenomenon later. All in all, the present results show that the variational method and the TRWA treatment are applicable to the strong-coupling regime when $\alpha\sim0.1$. This is a significant improvement over the SP treatment which is justified when $\alpha\leq0.01$.

Next, we focus on the multi-$D_1$ and TRWA results to analyze the emission spectral features of the two distant qubits. Figure~\ref{fig1}(a) shows that when $d=L_0$, the spectrum exhibits one narrow peak and one broad peak for the factorized initial state $\Psi_0$. In contrast, Figs.~\ref{fig1}(b)-\ref{fig1}(c) show that the spectrum exhibits a single broad peak for the correlated initial state $\Psi_{+}$ and exhibits a single narrow peak for the correlated initial state $\Psi_{-}$, corresponding to the superradiant and subradiant state, repsectively. Intuitively, it seems like that the formation of the doublet spectrum can be ascribed to the spontaneous emission from the correlated initial states. This situation is in analogy with the vacuum Rabi splitting which occurs in a qubit inside a cavity~\cite{Bishop_2008}. By using the analytical theory, we can figure out that the splitting between the two peaks are approximately given by $2|V_{c}+\tilde{\Delta}(\eta\omega_0,d)|$ when $d\sim L_0$, which depends on the reservoir induced dipole-dipole coupling strength $V_c$ and the Lamb shift $\tilde{\Delta}(\eta\omega_0,d)$.

Figures~\ref{fig1}(d)-\ref{fig1}(f) show that when $d=3L_{0}$, the spectrum exhibits a single peak for the three kinds of initial states. Moreover, the formation of the spectrum in~Fig.~\ref{fig1}(d) can be understood as a result of the superposition of the spectrum in Figs.~\ref{fig1}(e) and \ref{fig1}(f). It is worthwhile to note that the emission peak for the symmetric correlated initial state is very narrow, which indicates the occurrence of the subradiance. Figures~\ref{fig1}(g)-\ref{fig1}(i) show that when the distance $d$ further increases to $12L_{0}$, the spectrum is generally multipeaked for the three kinds of the initial states. The generation of the multipeaked spectrum is a result of the quantum interference of the radiating fields from the two distant qubits~\cite{Sinha_2020}. Besides, for large distances, the emission lines are in general sharply different from the typical Lorentzian lines, suggesting the nonMarkovian nature of the spontaneous emission process arising from the finite distance between the two qubits.

From Fig.~\ref{fig2} we see that the spectral features in the case of $\alpha=0.1$ are overall similar to those in the case of $\alpha=0.05$. Besides, Fig.~\ref{fig2} further confirms that the spectra from the factorized initial state are composed of the emission lines arising from the two correlated states. Interestingly, we observe the ultranarrow emission lines in Figs.~\ref{fig2}(g) and~\ref{fig2}(h), signifying the subradiance. The present results suggest that the emission spectrum strongly depends on the distance of the two qubits, and the subradiance and superradiance may occur by tuning the distance. In addition, we find that the spectrum are generally asymmetric irrespective of the distance in the strong-coupling regime, in contrast with the RWA symmetric spectra in the weak-coupling regime~\cite{Sinha_2020}.

We come to discuss in detail the subradiance in Fig.~\ref{fig2}(h) , which results in the ultranarrow emission line in the TRWA spectrum as well as the inconsistency between the multi-$D_1$ and TRWA spectra in Figs.~\ref{fig2}(g) and \ref{fig2}(h). To illustrate the subradiance, we employ the variational approach to calculate the dynamics of the excited-state population of the qubits as well as the dynamics of the photon  number for $\alpha=0.1$, $d=12L_{0}$, and the symmetric correlated initial state. Figure~\ref{fig3}(a) displays that the excited-state population of the two qubits oscillates at short times due to the dipole-dipole coupling, and as the time goes on this oscillation dies out and the population ultraslowly decays with vanishingly small beat behavior. Clearly, the qubits do not reach the steady state when $t=300\omega_0^{-1}$. Figure~\ref{fig3}(b) displays that the photon number at $\omega_k=0.8084\omega_0$ slowly increases with time while the photon number at $\omega_k=1.1250\omega_0$ oscillates with an ultraslowly decaying amplitude, indicating that it takes a long time for the field to arrive at the steady state. The present finding suggests that even in the presence of a strongly dissipative reservoir, the spontaneous decay of the two qubits can be ultraslow due to the subradiance, which is in sharply contrast to the case of a single qubit strongly coupled to the reservoir.

\section{Conclusions}\label{sec:con}
In summary, we have studied the emission spectrum of the two distant qubits strongly coupled to an Ohmic waveguide by using the variational approach and two analytical methods: TRWA and SP. The variational approach is based on the combination of the Dirac-Frenkel time-dependent variational principle and the multi-$D_{1}$ ansatz. The TRWA (SP) approach combines the resolvent-operator formalism and the TRWA (original) Hamiltonian, which allows us to derive analytical spectrum function. The variational approach and the TRWA treatment are found to be consistent with each other and valid in certain strong-coupling regimes where the SP treatment breaks down. By using the variational and TRWA approaches we have illustrated that the emission spectrum of the two distant qubits are generally asymmetric and can exhibits a single peak, double peaks, and multiple peaks depending on the distance and initial state of the two qubits. We also elucidate that in spite of the strong coupling between the qubits and the reservoir, the occurrence of the subradiance leads to that the two qubits and the radiation field ultraslowly reach their steady states. Our results provide insights into the emission spectral features of the two distant qubits in the strong light-matter coupling regime.

The variational approach with the multi-$D_1$ ansatz allows us to go beyond the widely used RWA and the Born-Markovian approximation. This approach captures not only the reduced dynamics but also the field dynamics in a single simulation and can be further extended and applied to waveguide QED problems involving few multi-level emitters in strong light-matter coupling regimes.

\begin{acknowledgments}
Support from the National Natural Science Foundation of China (Grants No.~12005188 and No.~11774311)
is gratefully acknowledged.
\end{acknowledgments}

\appendix

\section{Equations of motion for the variational parameters}\label{app:EOM}

\begin{figure*}
  \includegraphics[width=2\columnwidth]{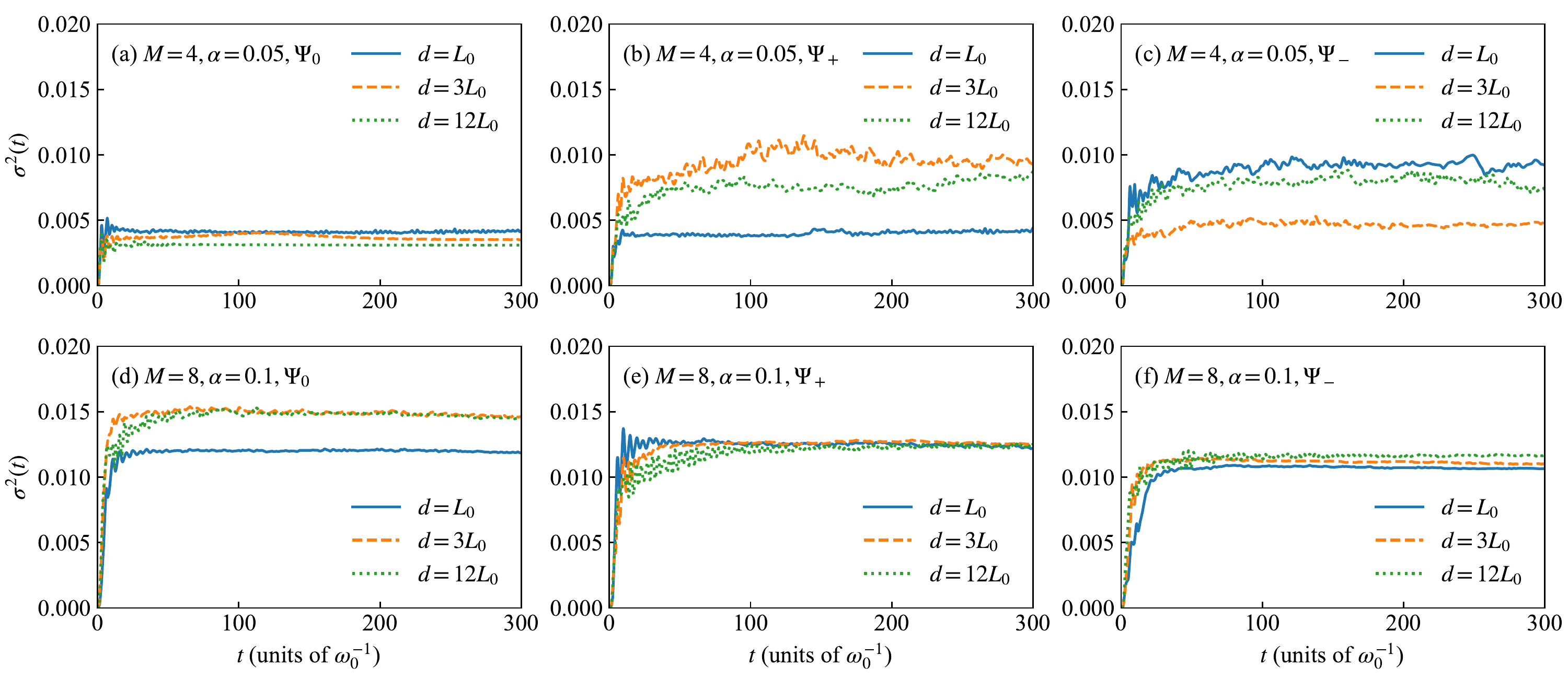}
  \caption{Scaled squared norm $\sigma^{2}(t)$ of the deviation vector as a function time $t$ for two values of $\alpha$, the three values of $d$, and the three kinds of initial states.}\label{fig4}
\end{figure*}

To obtain the equations of motion for the variational parameters, we differentiate the multi-$D_{1}$ state with respect to $t$, which yields
\begin{equation}
|\dot{D}_{1}^{M}(t)\rangle=\sum_{j=1}^{4}\sum_{n=1}^{M}\left(a_{nj}+A_{nj}\sum_{k}\dot{f}_{njk}b_{k}^{\dagger}\right)|\phi_{j}\rangle|f_{nj}\rangle,
\end{equation}
where
\begin{equation}
a_{nj}=\dot{A}_{nj}-\frac{1}{2}A_{nj}\sum_{k}(\dot{f}_{njk}f_{njk}^{\ast}+f_{njk}\dot{f}_{njk}^{\ast}). \label{eq:ani}
\end{equation}
By using the explicit form of the multi-$D_1$ state and its time derivative, it is straightforward to write the equations of motion~\eqref{eq:eom1} and~\eqref{eq:eom2} in terms of the variational parameters  as follows:
\begin{widetext}
\begin{eqnarray}
0 & = & i\sum_{n=1}^{M}\left(a_{nj}+A_{nj}\sum_{k}f_{mjk}^{\ast}\dot{f}_{njk}\right)\langle f_{mj}|f_{nj}\rangle-\sum_{l=1}^{4}\sum_{n=1}^{M}\langle\phi_{j}|H_{S}|\phi_{l}\rangle A_{nl}\langle f_{mj}|f_{nl}\rangle\nonumber \\
 &  & -\sum_{n=1}^{M}A_{nj}\sum_{h=1}^{2}\langle\phi_{j}|\sigma_{h}^{x}|\phi_{j}\rangle\sum_{k}\frac{\lambda_{k}}{2}(f_{mjk}^{\ast}e^{ikx_{h}+i\omega_{k}t}+f_{njk}e^{-ikx_{h}-i\omega_{k}t})\langle f_{mj}|f_{nj}\rangle.
\end{eqnarray}

\begin{eqnarray}
0 & = & i\sum_{n=1}^{M}\left(\sum_{j=1}^{4}A_{mj}^{\ast}a_{nj}f_{njq}+\sum_{j=1}^{4}A_{mj}^{\ast}A_{nj}\sum_{k}(\delta_{k,q}+f_{mjk}^{\ast}f_{njq})\dot{f}_{njk}\right)\langle f_{mj}|f_{nj}\rangle\nonumber \\
 &  & -\sum_{j,l=1}^{4}\sum_{n=1}^{M}A_{mj}^{\ast}A_{nl}\langle\phi_{j}|H_{S}|\phi_{l}\rangle\langle f_{mj}|f_{nl}\rangle-\sum_{j=1}^{4}\sum_{n=1}^{M}A_{mj}^{\ast}A_{nj}\sum_{h=1}^{2}\langle\phi_{j}|\sigma_{h}^{x}|\phi_{j}\rangle\nonumber \\
 &  & \times\sum_{k}\frac{\lambda_{k}}{2}\left[f_{njk}f_{njq}e^{-ikx_{h}-i\omega_{k}t}+(\delta_{k,q}+f_{mjk}^{\ast}f_{njq})e^{ikx_{h}+i\omega_{k}t}\right]\langle f_{mj}|f_{nj}\rangle.
\end{eqnarray}

These differential equations can be integrated by the following steps. First, we rewrite the above equations in a matrix form $i{\cal M}\vec{y}={\cal \vec{I}},$
where ${\cal M}$ is the coefficient matrix, $\vec{y}$ is a vector
consisting of $a_{nj}$ and $\dot{f}_{njk}$, and ${\cal \vec{I}}$
is the imhomogeneous term. Second, the matrix equation is solved to yield the values of $a_{nj}$ and $\dot{f}_{njk}$. $\dot{A}_{nj}$ is obtained via Eq.~\eqref{eq:ani}. Third, the values of the derivatives are used to update the variational parameters based on the 4th-order Runge-Kutta algorithm.

The scaled squared norm of the deviation vector depends on the squared norm of $|\dot{D}^M_1(t)\rangle$ and the mean value of $\tilde{H}^2(t)$ over the multi-$D_1$ state, which can be formally calculated as follows:
\begin{eqnarray}
\langle\dot{D}_{1}^{M}(t)|\dot{D}_{1}^{M}(t)\rangle & = & \sum_{j=1}^{4}\sum_{n,l=1}^{M}\left[a_{mj}^{\ast}a_{nj}+a_{mj}^{\ast}A_{nj}\sum_{k}\dot{f}_{njk}f_{mjk}^{\ast}+A_{mj}^{\ast}a_{nj}\sum_{k}\dot{f}_{mjk}^{\ast}f_{njk}\right.\nonumber \\
 &  & \left.+A_{mj}^{\ast}A_{nj}\sum_{k,q}(\delta_{k,q}+f_{mjk}^{\ast}f_{njq})\dot{f}_{mjq}^{\ast}\dot{f}_{njk}\right]\langle f_{mj}|f_{nj}\rangle,
\end{eqnarray}
\begin{eqnarray}
\left\langle \tilde{H}^{2}(t)\right\rangle  & = & \langle D_{1}^{M}(t)|\tilde{H}^{2}(t)|D_{1}^{M}(t)\rangle\nonumber \\
 & = & \sum_{j,l=1}^{4}\sum_{n,m=1}^{M}A_{mj}^{\ast}A_{nl}\langle\phi_{j}|H_{S}^{2}|\phi_{l}\rangle\langle f_{mj}|f_{nl}\rangle\nonumber \\
 &  & +\sum_{j,l=1}^{4}\sum_{n,m=1}^{M}\sum_{h=1}^{2}A_{mj}^{\ast}\langle\phi_{j}|\{H_{S},\sigma_{h}^{x}\}|\phi_{l}\rangle A_{nl}\sum_{k}\frac{\lambda_{k}}{2}(f_{mjk}^{\ast}e^{ikx_{h}+i\omega_{k}t}+f_{nlk}e^{-ikx_{h}-i\omega_{k}t})\langle f_{mj}|f_{nl}\rangle\nonumber \\
 &  & +\sum_{j=1}^{4}\sum_{n,m=1}^{M}\sum_{h=1}^{2}A_{mj}^{\ast}A_{nj}\left\{ \left[\sum_{k}\frac{\lambda_{k}}{2}(f_{mjk}^{\ast}e^{ikx_{h}+i\omega_{k}t}+f_{njk}e^{-ikx_{h}-i\omega_{k}t})\right]^{2}+\sum_{k}\frac{\lambda_{k}^{2}}{4}\right\} \langle f_{mj}|f_{nj}\rangle\nonumber \\
 &  & +2\sum_{j=1}^{4}\sum_{n,m=1}^{M}A_{mj}^{\ast}A_{nj}\langle\phi_{j}|\sigma_{1}^{x}\sigma_{2}^{x}|\phi_{j}\rangle\left[\sum_{k}\frac{\lambda_{k}}{2}(f_{mjk}^{\ast}e^{ikx_{1}+i\omega_{k}t}+f_{njk}e^{-ikx_{1}-i\omega_{k}t})\right.\nonumber \\
 &  & \left.\times\sum_{q}\frac{\lambda_{q}}{2}(f_{mjq}^{\ast}e^{iqx_{2}+i\omega_{q}t}+f_{njq}e^{-iqx_{2}-i\omega_{q}t})+\sum_{k}\frac{\lambda_{k}^{2}}{4}\cos(kd)\right]\langle f_{mj}|f_{nj}\rangle.
\end{eqnarray}
\end{widetext}
In Fig.~\ref{fig4}, we plot the behaviors of $\sigma^2(t)$ versus time for the two values of $\alpha$, the three values of $d$, and the three kinds of the initial states. It turns out that the magnitude of $\sigma^{2}(t)$ takes on relatively small values, the order of which is $10^{-2}$ or smaller. This guarantees the reliability of the variational results~\cite{Yan_2020}.

\section{Derivation of the effective Hamiltonian}\label{app:Heff}
The Hamiltonian in the transformed frame can be readily derived as follows:
\begin{eqnarray}
H^{\prime} & = & e^{S}He^{-S}\nonumber \\
 & = & \frac{\omega_{0}}{2}\sum_{j=1}^{2}\left(\sigma_{j}^{z}\cosh X_{j}-i\sigma_{j}^{y}\sinh X_{j}\right)\nonumber \\
 &  & +\sum_{k}\omega_{k}b_{k}^{\dagger}b_{k}+\sum_{k}\frac{\lambda_{k}^{2}}{2\omega_{k}}(\xi_{k}^{2}-2\xi_{k})\nonumber \\
 &  & +\sum_{j=1}^{2}\frac{\sigma_{j}^{x}}{2}\sum_{k}\lambda_{k}(1-\xi_{k})\left(b_{k}e^{-ikx_{j}}+b_{k}^{\dagger}e^{ikx_{j}}\right)\nonumber \\
 &  & +\sum_{k}\frac{\lambda_{k}^{2}}{2\omega_{k}}(\xi_{k}^{2}-2\xi_{k})\cos\left(kd\right)\sigma_{1}^{x}\sigma_{2}^{x},
\end{eqnarray}
where
\begin{equation}
X_{j}=\frac{\lambda_{k}}{\omega_{k}}\xi_{k}(b_{k}^{\dagger}e^{ikx_{j}}-b_{k}e^{-ikx_{j}}).
\end{equation}
We divide the transformed Hamiltonian into three parts:
\begin{equation}
H^{\prime}=H_{0}^{\prime}+H_{1}^{\prime}+H_{2}^{\prime},
\end{equation}
\begin{equation}
H_{0}^{\prime}=\frac{1}{2}\eta\omega_{0}\sum_{j=1}^{2}\sigma_{j}^{z}+\sum_{k}\omega_{k}b_{k}^{\dagger}b_{k}+\sum_{k}\frac{\lambda_{k}^{2}}{2\omega_{k}}(\xi_{k}^{2}-2\xi_{k}),
\end{equation}
\begin{eqnarray}
H_{1}^{\prime} & = & \sum_{j=1}^{2}\frac{\sigma_{j}^{x}}{2}\sum_{k}\lambda_{k}(1-\xi_{k})\left(b_{k}e^{-ikx_{j}}+b_{k}^{\dagger}e^{ikx_{j}}\right)\nonumber \\
 &  & -i\eta\omega_{0}\sum_{j=1}^{2}\frac{\sigma_{j}^{y}}{2}\sum_{k}\frac{\lambda_{k}}{\omega_{k}}\xi_{k}\left(b_{k}^{\dagger}e^{ikx_{j}}-b_{k}e^{-ikx_{j}}\right)\nonumber \\
 &  &+V_{c}\sigma_{1}^{x}\sigma_{2}^{x},
\end{eqnarray}
\begin{equation}
H_{2}^{\prime}=\frac{\omega_{0}}{2}\sum_{j=1}^{2}(\cosh X_{j}-\eta)\sigma_{j}^{z}-i\frac{\omega_{0}}{2}\sum_{j=1}^{2}(\sinh X_{j}-\eta X_{j})\sigma_{j}^{y},
\end{equation}
where
\begin{equation}
\eta=\langle0|\cosh X_{j}|0\rangle=\exp\left(-\frac{1}{2}\sum_{k}\frac{\lambda_{k}^{2}}{\omega_{k}^{2}}\xi_{k}^{2}\right),
\end{equation}
\begin{equation}
V_{c}=\sum_{k}\frac{\lambda_{k}^{2}}{2\omega_{k}}(\xi_{k}^{2}-2\xi_{k})\cos(kd).
\end{equation}

To proceed, we reformulate the qubit-reservoir interaction in $H_{1}^{\prime}$
to be the RWA form, which can be achieved by setting
\begin{equation}
\lambda_{k}(1-\xi_{k})=\eta\omega_{0}\frac{\lambda_{k}}{\omega_{k}}\xi_{k}.
\end{equation}
This equation results in
$\xi_{k}=\frac{\omega_{k}}{\omega_{k}+\eta\omega_{0}}$
and
$H_{1}^{\prime}=V_{c}\sigma_{1}^{x}\sigma_{2}^{x}+\sum_{j=1}^{2}\sum_{k}\tilde{\lambda}_{k}\left(\sigma_{j}^{+}b_{k}e^{-ikx_{j}}+\sigma_{j}^{-}b_{k}^{\dagger}e^{ikx_{j}}\right)$.
We should emphasis that the value of $\xi_k$ can also be obtained by minimizing the ground-state energy of $H_0^\prime$. Up till now we have not introduced any approximations.

To make analytical calculation managable, we use $H^{\prime}\approx H_{0}^{\prime}+H_{1}^{\prime}$
as the effective Hamiltonian because $H_{2}^{\prime}$ comprises of
the second- and higher-order bosonic processes, the contribution of which is the order of $\lambda_k^4$. This approximation is expected to
be reasonable in a moderately strong coupling regime and is referred to as the TRWA.

\section{Analytical calculation of spontaneous emission spectrum}\label{app:sw}

Without loss of generality, we consider the initial state $|eg0\rangle\equiv|eg\rangle\otimes|0\rangle$. In the laboratory frame, the transition amplitude associated with the spontaneous emission
process is given by
\begin{equation}
\langle gg1_{k}|U(t)|eg0\rangle,
\end{equation}
where $|gg1_k\rangle$ denotes the state that the two qubits are in the ground state and one photon occupies the $k$th mode of the reservoir, and $U(t)=\exp(-iHt)$ is the time-evolution operator of the whole
system. The steady photon number at the $k$th mode is then calculated
as
\begin{equation}
N(k)=\lim_{t\rightarrow\infty}|\langle gg1_{k}|U(t)|eg0\rangle|^{2}.
\end{equation}

We use the resolvent-operator formalism to calculate the transition amplitude. This formalism relates
 the time-evolution operator to the resolvent operator via the integral~\cite{Cohen},
\begin{equation}
U(t)=\frac{1}{2\pi i}\int_{+\infty}^{-\infty}G(\omega+i0^{+})e^{-i\omega t}d\omega,\quad(t>0)
\end{equation}
where
\begin{equation}
G(z)=\frac{1}{z-H} \label{eq:Gz}
\end{equation}
is the resolvent operator and $z$ is a complex variable. In general, $H$ is able to be divided into two parts: an exactly diagonalized $H_0$ and a perturbation $V$.

To calculate the matrix elements of the resolvent operator between some bases, e.g., the interested eigenstates of $H_0$, we introduce
projectors ${\cal P}$ and ${\cal Q}=1-{\cal P}$, where ${\cal P}$ projects onto a subspace spanned by some interested eigenstates of $H_0$ and ${\cal Q}$ projects onto the complementary space.
By using these projectors, we can derive the following equations from Eq.~\eqref{eq:Gz},
\begin{equation}
{\cal P}(z-H){\cal P}G(z){\cal P}-{\cal P}V{\cal Q}G(z){\cal P}={\cal P},\label{eq:PGeq}
\end{equation}
\begin{equation}
{\cal Q}(z-H){\cal Q}G(z){\cal P}-{\cal Q}V{\cal P}G(z){\cal P}=0.
\end{equation}
The second equation can be solved to yield
\begin{equation}
{\cal Q}G(z){\cal P}=\frac{1}{{\cal Q}(z-H){\cal Q}}{\cal Q}V{\cal P}G(z){\cal P}. \label{eq:QGP0}
\end{equation}
Substituting this solution of ${\cal Q}G(z){\cal P}$ into Eq.~\eqref{eq:PGeq}, one readily obtains
\begin{equation}
{\cal P}G(z){\cal P}=\frac{{\cal P}}{z-H_{0}-{\cal P}R(z){\cal P}}, \label{eq:PGP}
\end{equation}
where
\begin{eqnarray}
R(z) & = & V+V\frac{{\cal Q}}{{\cal Q}(z-H){\cal Q}}V\nonumber \\
 & = & V+V\frac{{\cal Q}}{z-H_{0}}V+\ldots \label{eq:Rz}
\end{eqnarray}
is the level-shift operator. Using Eq.~\eqref{eq:PGP}, we can rewrite ${\cal Q}G(z){\cal P}$ as
\begin{equation}
{\cal Q}G(z){\cal P}=\frac{1}{{\cal Q}(z-H){\cal Q}}{\cal Q}V\frac{{\cal P}}{z-H_{0}-{\cal P}R(z){\cal P}}.\label{eq:QGP}
\end{equation}

Similarly, one can derive the expression of ${\cal Q}G(z){\cal Q}$ as follows:
\begin{eqnarray}
{\cal Q}G(z){\cal Q} & = & \frac{{\cal Q}}{z-{\cal Q}H{\cal Q}}+\frac{{\cal Q}}{z-{\cal Q}H{\cal Q}}V\nonumber \\
 &  & \times\frac{{\cal P}}{z-H_{0}-{\cal P}R(z){\cal P}}V\frac{{\cal Q}}{z-{\cal Q}H{\cal Q}}. \label{eq:QGQ}
\end{eqnarray}

In the following, we show that the transition amplitudes associated with the spontaneous emission process can be derived from Eqs.~\eqref{eq:PGP},~\eqref{eq:QGP}, and~\eqref{eq:QGQ}.

\subsection{Spontaneous emission spectrum with the unitary transformation}

In this section, we calculate the transition amplitude with the effective Hamiltonian in the transformed frame. With the unitary transformation, we have
\begin{widetext}
\begin{eqnarray}
\langle gg1_{k}|U(t)|eg0\rangle & = & \langle gg1_{k}|e^{-S}e^{S}U(t)e^{-S}e^{S}|eg0\rangle\nonumber \\
 & \approx & \langle gg1_{k}|U^{\prime}(t)|eg0\rangle+\sum_{q}\frac{\lambda_{q}}{2\omega_{q}}\xi_{q}e^{iqx_{1}}\langle gg1_{k}|U^{\prime}(t)|gg1_{q}\rangle\nonumber \\
 &  & -\frac{\lambda_{k}}{2\omega_{k}}\xi_{k}e^{ikx_{1}}\langle eg0|U^{\prime}(t)|eg0\rangle-\frac{\lambda_{k}}{2\omega_{k}}\xi_{k}e^{ikx_{2}}\langle ge0|U^{\prime}(t)|eg0\rangle, \label{eq:UTP}
\end{eqnarray}
\end{widetext}
where we have used $e^{S}U(t)e^{-S}\approx\exp(-iH^{\prime}t)\equiv U^{\prime}(t)$
and just retained the transition amplitudes associated with the single-excitation
states. It is evident that in the transformed frame we need to evaluate four transition amplitudes with the transformed Hamiltonian $H^\prime$.

Using ${\cal P}=|eg0\rangle\langle eg0|+|ge0\rangle\langle ge0|$ and setting $H_0^\prime$ and $H_1^\prime$ as the free and interaction Hamiltonian, respectively, we can derive the elements of the level-shift operator as follows:
\begin{equation}
\langle eg0|R(z)|eg0\rangle  =  \langle ge0|R(z)|ge0\rangle=\sum_{q}\frac{\tilde{\lambda}_{q}^{2}}{z+\eta\omega_{0}-\omega_{q}},
\end{equation}
\begin{equation}
  \langle ge0|R(z)|eg0\rangle=\langle eg0|R(z)|ge0\rangle=V_{c}+\sum_{q}\frac{\tilde{\lambda}_{q}^{2}\cos(qd)}{z+\eta\omega_{0}-\omega_{q}},
\end{equation}

It follows from Eq.~\eqref{eq:PGP} that
\begin{equation}
\langle eg0|G(z)|eg0\rangle=\langle ge0|G(z)|ge0\rangle=\frac{z-\sum_{q}\frac{\tilde{\lambda}_{q}^{2}}{z+\eta\omega_{0}-\omega_{q}}}{\tilde{D}(z)},
\end{equation}
\begin{equation}
\langle ge0|G(z)|eg0\rangle=\langle eg0|G(z)|ge0\rangle=\frac{V_{c}+\sum_{q}\frac{\tilde{\lambda}_{q}^{2}\cos(qd)}{z+\eta\omega_{0}-\omega_{q}}}{\tilde{D}(z)},
\end{equation}
where
\begin{eqnarray}
\tilde{D}(z) &=& \left[z-\sum_{q}\frac{\tilde{\lambda}_{q}^{2}}{z+\eta\omega_{0}-\omega_{q}}\right]^{2}\nonumber\\
& &-\left[V_{c}+\sum_{q}\frac{\tilde{\lambda}_{q}^{2}\cos(qd)}{z+\eta\omega_{0}-\omega_{q}}\right]^{2}.
\end{eqnarray}
Using Eq.~\eqref{eq:QGP} we find
\begin{widetext}
\begin{eqnarray}
\langle gg1_{k}|G(z)|eg0\rangle & = & \langle gg1_{k}|\frac{{\cal Q}}{{\cal Q}(z-H^{\prime}){\cal Q}}H_{1}^{\prime}{\cal P}G(z){\cal P}|eg0\rangle\nonumber\\
 & = & \langle gg1_{k}|\frac{{\cal Q}}{{\cal Q}(z-H^{\prime}){\cal Q}}H_{1}^{\prime}|eg0\rangle\langle eg0|G(z)|eg0\rangle +\langle gg1_{k}|\frac{{\cal Q}}{{\cal Q}(z-H^{\prime}){\cal Q}}H_{1}^{\prime}|ge0\rangle\langle ge0|G(z)|eg0\rangle\nonumber\\
 & \approx & \tilde{\lambda}_{k}e^{ikx_{1}}\langle gg1_{k}|\frac{{\cal Q}}{{\cal Q}(z-H^{\prime}){\cal Q}}|gg1_{k}\rangle[\langle eg0|G(z)|eg0\rangle+e^{-ikd}\langle ge0|G(z)|eg0\rangle]\nonumber\\
 & = & \tilde{\lambda}_{k}e^{ikx_{1}}\frac{1}{z+\eta\omega_{0}-\omega_{k}-\frac{V_{c}^{2}}{z-\eta\omega_{0}-\omega_{k}}}\left[\frac{z-\sum_{q}\frac{\tilde{\lambda}_{q}^{2}}{z+\eta\omega_{0}-\omega_{q}}}{\tilde{D}(z)}+e^{-ikd}\frac{V_{c}+\sum_{q}\frac{\tilde{\lambda}_{q}^{2}\cos(qd)}{z+\eta\omega_{0}-\omega_{q}}}{\tilde{D}(z)}\right],
\end{eqnarray}
\end{widetext}
where we have used
\begin{equation}
\langle gg1_{k}|\frac{{\cal Q}}{z-{\cal Q}H^{\prime}{\cal Q}}|gg1_{q}\rangle\approx\frac{\delta_{k,q}}{z+\eta\omega_{0}-\omega_{k}-\frac{V_{c}^{2}}{z-\eta\omega_{0}-\omega_{k}}},
\end{equation}
which is derived from Eq.~\eqref{eq:QGQ}.

To proceed, we should replace $z$ with $\omega+i0^{+}$ in the matrix elements of the resolvent operator before integrating and use
\begin{equation}
\sum_{q}\frac{\tilde{\lambda}_{q}^{2}\cos(qd)}{\omega-\omega_{q}+i0^{+}}=\tilde{\Delta}\left(\omega,d\right)-i\tilde{\Gamma}\left(\omega,d\right),
\end{equation}
where
\begin{equation}
\tilde{\Delta}\left(\omega,d\right)=P\sum_{q}\frac{\tilde{\lambda}_{q}^{2}\cos(qd)}{\omega-\omega_{q}},
\end{equation}
\begin{equation}
\tilde{\Gamma}(\omega,d)=\pi\sum_{k}\tilde{\lambda}_{q}^{2}\cos(qd)\delta(\omega_{q}-\omega).
\end{equation}

In the long-time limit $t\rightarrow\infty$ and $d\neq0$, the transition amplitudes given in Eq.~\eqref{eq:UTP} can be evaluated as follows:
\begin{widetext}
\begin{equation}
\langle eg0|U^{\prime}(t)|eg0\rangle  =  \langle ge0|U^{\prime}(t)|ge0\rangle =\frac{1}{2\pi i}\int_{\infty}^{-\infty}d\omega e^{-i\omega t}\frac{\tilde{A}(\omega+\eta\omega_{0})}{\tilde{A}^{2}(\omega+\eta\omega_{0})-\tilde{B}^{2}(\omega+\eta\omega_{0})}=0,
\end{equation}
\begin{equation}
\langle ge0|U^{\prime}(t)|eg0\rangle  =  \langle eg0|U^{\prime}(t)|ge0\rangle
  =  \frac{1}{2\pi i}\int_{\infty}^{-\infty}d\omega e^{-i\omega t}\frac{\tilde{B}(\omega+\eta\omega_{0})}{\tilde{A}^{2}(\omega+\eta\omega_{0})-\tilde{B}^{2}(\omega+\eta\omega_{0})}=0,
\end{equation}
\begin{eqnarray}
\langle gg1_{k}|U^{\prime}(t)|gg1_{p}\rangle & \approx & \frac{1}{2\pi i}\int_{\infty}^{-\infty}d\omega e^{-i\omega t}\delta_{k,q}\frac{1}{\omega+\eta\omega_{0}-\omega_{k}-\frac{V_{c}^{2}}{\omega-\eta\omega_{0}-\omega_{k}}} \nonumber\\
&=&  \delta_{k,q}\exp\left[-i\left(\omega_{k}-\eta\omega_{0}-\frac{V_{c}^{2}}{2\eta\omega_{0}}\right)t\right],
\end{eqnarray}
\begin{eqnarray}
\langle gg1_{k}|U^{\prime}(t)|eg0\rangle & = & \frac{1}{2\pi i}\int_{\infty}^{-\infty}d\omega e^{-i\omega t}\tilde{\lambda}_{k}e^{ikx_{1}}\frac{1}{\omega+\eta\omega_{0}-\omega_{k}-\frac{V_{c}^{2}}{\omega-\eta\omega_{0}-\omega_{k}}}\frac{\tilde{A}(\omega+\eta\omega_{0})+e^{-ikd}\tilde{B}(\omega+\eta\omega_{0})}{\tilde{A}^{2}(\omega+\eta\omega_{0})-\tilde{B}^{2}(\omega+\eta\omega_{0})}\nonumber \\
 & = & \tilde{\lambda}_{k}e^{ikx_{1}}\frac{\tilde{A}\left(\omega_{k}-\frac{V_{c}^{2}}{2\eta\omega_{0}}\right)+e^{-ikd}\tilde{B}\left(\omega_{k}-\frac{V_{c}^{2}}{2\eta\omega_{0}}\right)}{\tilde{A}^{2}\left(\omega_{k}-\frac{V_{c}^{2}}{2\eta\omega_{0}}\right)-\tilde{B}^{2}\left(\omega_{k}-\frac{V_{c}^{2}}{2\eta\omega_{0}}\right)}\exp\left[-i\left(\omega_{k}-\eta\omega_{0}-\frac{V_{c}^{2}}{2\eta\omega_{0}}\right)t\right],
\end{eqnarray}
where we have used the fact that in the long-time limit the simple pole $\omega\approx\omega_k-\frac{V_{c}^{2}}{2\eta\omega_0}$ contributes to the steady state. $\tilde{A}(\omega)$ and $\tilde{B}(\omega)$ are defined in Eqs.~\eqref{eq:Awt} and ~\eqref{eq:Bwt} in the main text.
Similarly, we also have
\begin{equation}
\langle gg1_{k}|U^{\prime}(t)|ge0\rangle=\tilde{\lambda}_{k}e^{ikx_{1}}\frac{e^{-ikd}\tilde{A}\left(\omega_{k}-\frac{V_{c}^{2}}{2\eta\omega_{0}}\right)+\tilde{B}\left(\omega_{k}-\frac{V_{c}^{2}}{2\eta\omega_{0}}\right)}{\tilde{A}^{2}\left(\omega_{k}-\frac{V_{c}^{2}}{2\eta\omega_{0}}\right)-\tilde{B}^{2}\left(\omega_{k}-\frac{V_{c}^{2}}{2\eta\omega_{0}}\right)}\exp\left[-i\left(\omega_{k}-\eta\omega_{0}-\frac{V_{c}^{2}}{2\eta\omega_{0}}\right)t\right].
\end{equation}
\end{widetext}
Using above transition amplitudes, we can calculate the photon number at the $k$th mode in the long-time limit and thus derive the steady-state emission spectra.

\subsection{Standard perturbation calculation}
In this section, we calculate the spontaneous emission spectrum with the original Hamiltonian by setting $H_0=\frac{\omega_{0}}{2}\sum_{j=1}^{2}\sigma_{j}^{z}+\sum_{k}\omega_{k}b_{k}^{\dagger}b_{k}$ and $V=\sum_{j=1}^{2}\frac{\sigma_{j}^{x}}{2}\sum_{k}\lambda_{k}\left(b_{k}e^{-ikx_{j}}+b_{k}^{\dagger}e^{ikx_{j}}\right)$. Similarly, we use
${\cal P}=|eg0\rangle\langle eg0|+|ge0\rangle\langle ge0|$. The matrix elements of the level-shift operator can be computed up to the second order in $\lambda_k$ as follows:
\begin{eqnarray}
  \langle eg0|R(z)|eg0\rangle& = & \langle ge0|R(z)|ge0\rangle\nonumber\\
  & =& \sum_{q}\left(\frac{\lambda_{q}^{2}/4}{z+\omega_{0}-\omega_{q}}+\frac{\lambda_{q}^{2}/4}{z-\omega_{0}-\omega_{q}}\right),\nonumber\\
\end{eqnarray}
\begin{eqnarray}
  \langle ge0|R(z)|eg0\rangle& = &\langle ge0|R(z)|eg0\rangle\nonumber\\
  & =& \sum_{q}\left(\frac{\lambda_{q}^{2}\cos(qd)/4}{z+\omega_{0}-\omega_{q}}+\frac{\lambda_{q}^{2}\cos(qd)/4}{z-\omega_{0}-\omega_{q}}\right).\nonumber\\
\end{eqnarray}

From Eq.~\eqref{eq:PGP} one readily derives the following matrix elements:
\begin{equation}
\langle eg0|G(z)|eg0\rangle=\frac{z-\sum_{q}\left(\frac{\lambda_{q}^{2}\cos(qd)/4}{z+\omega_{0}-\omega_{q}}+\frac{\lambda_{q}^{2}\cos(qd)/4}{z-\omega_{0}-\omega_{q}}\right)}{D(z)},
\end{equation}
\begin{equation}
\langle ge0|G(z)|eg0\rangle=\frac{\sum_{q}\left(\frac{\lambda_{q}^{2}\cos(qd)/4}{z+\omega_{0}-\omega_{q}}+\frac{\lambda_{q}^{2}\cos(qd)/4}{z-\omega_{0}-\omega_{q}}\right)}{D(z)},
\end{equation}
where
\begin{eqnarray}
D(z) & = & \left[z-\sum_{q}\left(\frac{\lambda_{q}^{2}/4}{z+\omega_{0}-\omega_{q}}+\frac{\lambda_{q}^{2}/4}{z-\omega_{0}-\omega_{q}}\right)\right]^{2}\nonumber \\
 &  & -\left[\sum_{q}\left(\frac{\lambda_{q}^{2}\cos(qd)/4}{z+\omega_{0}-\omega_{q}}+\frac{\lambda_{q}^{2}\cos(qd)/4}{z-\omega_{0}-\omega_{q}}\right)\right]^{2}.\nonumber\\
\end{eqnarray}

Using Eq.~\eqref{eq:QGP}, we have
\begin{widetext}
\begin{eqnarray}
\langle gg1_{k}|G(z)|eg0\rangle & = & \langle gg1_{k}|\frac{{\cal Q}}{{\cal Q}(z-H){\cal Q}}V{\cal P}G(z)|eg0\rangle\nonumber \\
 & = & \langle gg1_{k}|\frac{{\cal Q}}{{\cal Q}(z-H){\cal Q}}V|eg0_{k}\rangle\langle eg0|G(z)|eg0\rangle +\langle gg1_{k}|\frac{{\cal Q}}{{\cal Q}(z-H){\cal Q}}V|ge0_{k}\rangle\langle ge0|G(z)|eg0\rangle\nonumber \\
 & \approx & \frac{\lambda_{k}}{2}e^{ikx_{1}}\langle gg1_{k}|\frac{{\cal Q}}{{\cal Q}(z-H){\cal Q}}|gg1_{k}\rangle\langle eg0|G(z)|eg0\rangle +\frac{\lambda_{k}}{2}e^{ikx_{2}}\langle gg1_{k}|\frac{{\cal Q}}{{\cal Q}(z-H){\cal Q}}|gg1_{k}\rangle\langle ge0|G(z)|eg0\rangle\nonumber \\
 & = & \frac{\lambda_{k}}{2}e^{ikx_{1}}\langle gg1_{k}|\frac{{\cal Q}}{{\cal Q}(z-H){\cal Q}}|gg1_{k}\rangle[\langle eg0|G(z)|eg0\rangle +e^{-ikd}\langle ge0|G(z)|eg0\rangle]\nonumber\\
 & = & \frac{\lambda_{k}}{2}e^{ikx_{1}}\frac{1}{z+\omega_{0}-\omega_{k}-\frac{1}{2}\sum_{q}\frac{\lambda_{q}^{2}}{z-\omega_{k}-\omega_{q}}}\\
 &  & \times \frac{z-\sum_{q}\left(\frac{\lambda_{q}^{2}/4}{z+\omega_{0}-\omega_{q}}+\frac{\lambda_{q}^{2}/4}{z-\omega_{0}-\omega_{q}}\right)+e^{-ikd}\sum_{q}\left(\frac{\lambda_{q}^{2}\cos(qd)/4}{z+\omega_{0}-\omega_{q}}+\frac{\lambda_{q}^{2}\cos(qd)/4}{z-\omega_{0}-\omega_{q}}\right)}{D(z)} ,
\end{eqnarray}
where we have used
\begin{equation}
\langle gg1_{k}|\frac{{\cal Q}}{{\cal Q}(z-H){\cal Q}}|gg1_{k}\rangle\approx\frac{1}{z+\omega_{0}-\omega_{k}-\frac{1}{2}\sum_{q}\frac{\lambda_{q}^{2}}{z-\omega_{k}-\omega_{q}}}.
\end{equation}
Similarly, we have
\begin{eqnarray}
\langle gg1_{k}|G(z)|ge0\rangle & = & \frac{\lambda_{k}}{2}e^{ikx_{1}}\frac{1}{z+\omega_{0}-\omega_{k}-\frac{1}{2}\sum_{q}\frac{\lambda_{q}^{2}}{z-\omega_{k}-\omega_{q}}}\nonumber\\
 &  & \times \frac{e^{-ikd}\left[z-\sum_{q}\left(\frac{\lambda_{q}^{2}/4}{z+\omega_{0}-\omega_{q}}+\frac{\lambda_{q}^{2}/4}{z-\omega_{0}-\omega_{q}}\right)\right]+\sum_{q}\left(\frac{\lambda_{q}^{2}\cos(qd)/4}{z+\omega_{0}-\omega_{q}}+\frac{\lambda_{q}^{2}\cos(qd)/4}{z-\omega_{0}-\omega_{q}}\right)}{D(z)} ,
\end{eqnarray}

To perform the integral we
replace $z$ with $\omega+i0^{+}$ and use
\begin{equation}
\sum_{q}\frac{\lambda_{q}^{2}\cos(qd)/4}{\omega-\omega_{q}+i0^{+}}=\Delta(\omega,d)-i\Gamma(\omega,d),
\end{equation}
where
\begin{equation}
\Delta(\omega,d)=P\sum_{q}\frac{\lambda_{q}^{2}\cos(qd)/4}{\omega-\omega_{q}},
\end{equation}
\begin{equation}
\Gamma(\omega,d)=\frac{\pi}{4}\sum_{q}\lambda_{q}^{2}\cos(qd)\delta(\omega-\omega_{q}).
\end{equation}

In the long-time limit, we have
\begin{eqnarray}
\langle gg1_{k}|U(t)|eg0\rangle & = & \frac{1}{2\pi i}\int_{+\infty}^{-\infty}\langle gg1_{k}|G(\omega+i0^{+})|eg0\rangle e^{-i\omega t}d\omega\nonumber \\
 & = & \frac{1}{2\pi i}\int_{+\infty}^{-\infty}\frac{\lambda_{k}}{2}e^{ikx_{1}}\frac{1}{\omega+\omega_{0}-\omega_{k}-2\Delta_{0}(\omega-\omega_{k})}\times\frac{A(\omega+\omega_{0})+e^{-ikd}B(\omega+\omega_{0})}{A^{2}(\omega+\omega_{0})-B^{2}(\omega+\omega_{0})}e^{-i\omega t}d\omega\nonumber \\
 & \approx & \frac{\lambda_{k}}{2}e^{ikx_{1}}\frac{A[\omega_{k}+2\Delta(-\omega_{0},0)]+e^{-ikd}B[\omega_{k}+2\Delta(-\omega_{0},0)]}{A^{2}[\omega_{k}+2\Delta(-\omega_{0},0)]-B^{2}[\omega_{k}+2\Delta(-\omega_{0},0)]}e^{-i[\omega_{k}-\omega_{0}+2\Delta(-\omega_{0},0)]t},
\end{eqnarray}
\begin{eqnarray}
\langle gg1_{k}|U(t)|ge0\rangle & = & \frac{1}{2\pi i}\int_{+\infty}^{-\infty}\langle gg1_{k}|G(\omega+i0^{+})|ge0\rangle e^{-i\omega t}d\omega\nonumber \\
 & = & \frac{1}{2\pi i}\int_{+\infty}^{-\infty}d\omega e^{-i\omega t}\frac{\lambda_{k}}{2}e^{ikx_{1}}\frac{1}{\omega+\omega_{0}-\omega_{k}-2\Delta_{0}(\omega-\omega_{k})}\times\frac{e^{-ikd}A(\omega+\omega_{0})+B(\omega+\omega_{0})}{A^{2}(\omega+\omega_{0})-B^{2}(\omega+\omega_{0})}\nonumber \\
 & \approx & \frac{\lambda_{k}}{2}e^{ikx_{1}}\frac{e^{-ikd}A[\omega_{k}+2\Delta(-\omega_{0},0)]+B[\omega_{k}+2\Delta(-\omega_{0},0)]}{A^{2}[\omega_{k}+2\Delta(-\omega_{0},0)]-B^{2}[\omega_{k}+2\Delta(-\omega_{0},0)]}e^{-i[\omega_{k}-\omega_{0}+2\Delta(-\omega_{0},0)]t},
\end{eqnarray}
\end{widetext}
where $A(\omega)$ and $B(\omega)$ are defined in Eqs.~\eqref{eq:Aw} and \eqref{eq:Bw} in the main text, respectively, and we have used the fact that the simple pole $\omega\approx\omega_{k}-\omega_{0}+2\Delta(-\omega_{0},0)$ determined from $\omega+\omega_{0}-\omega_{k}-2\Delta(\omega-\omega_{k},0)=0$ contributes to the long-time behavior.
Using the above results, it is straightforward to calculate the emission spectra for the three kinds of the initial states.

\bibliography{twoatom}

\end{document}